\documentclass[fleqn,10pt]{wlscirep}
\usepackage[utf8]{inputenc}
\usepackage[T1]{fontenc}
\title{Typhoon track prediction using satellite images in a Generative Adversarial Network}

\author[1]{Mario Rüttgers}
\author[1]{Sangseung Lee}
\author[1,*]{Donghyun You}
\affil[1]{Pohang University of Science and Technology, Mechanical Engineering, Pohang, 37673, South-Korea}

\affil[*]{dhyou@postech.ac.kr}

\begin{abstract}
Tracks of typhoons are predicted using satellite images as input for a Generative Adversarial Network (GAN). The satellite images have time gaps of 6 hours and are marked with a red square at the location of the typhoon center. The GAN uses images from the past to generate an image one time step ahead. The generated image shows the future location of the typhoon center, as well as the future cloud structures. The errors between predicted and real typhoon centers are measured quantitatively in kilometers. $42.4 \%$ of all typhoon center predictions have absolute errors of less than $80$ km, $32.1 \%$ lie within a range of $80$ - $120$ km and the remaining $25.5 \%$ have accuracies above $120$ km. The relative error sets the above mentioned absolute error in relation to the distance that has been traveled by a typhoon over the past 6 hours. High relative errors are found in three types of situations, when a typhoon moves on the open sea far away from land, when a typhoon changes its course suddenly and when a typhoon is about to hit the mainland. The cloud structure prediction is evaluated qualitatively. It is shown that the GAN is able to predict trends in cloud motion. In order to improve both, the typhoon center and cloud motion prediction, the present study suggests to add information about the sea surface temperature, surface pressure and velocity fields to the input data.      
\end{abstract}
\begin{document}

\flushbottom
\maketitle
%
%
\thispagestyle{empty}

\section{Introduction}

Every year tropical cyclones cause death and damage in many places around the world. They are formed when water at the sea surface becomes warm, evaporates, rises in form of clouds and while the clouds cool down, the condensation releases strong energies in form of winds. The rotation of the earth gives cyclones its spinning motion. In the center usually a hole forms, which is called the eye of the cyclone. At the eye, the pressure is low and energetic clouds and winds get attracted. Warm air can rise through the hole and, favored by high altitude winds that can create a suction effect, increase the energy of the system. Crosswinds, on the other hand, can block this mechanism. Depending on the water conditions and surrounding winds, tropical cyclones and their associate storm surges can be a great danger when they find their way to populated land. 

The destructive force of a tropical cyclone comes from the wind speed of the rotating air and the rainfall that can cause disastrous flooding\cite{spiegel18}. The highest wind speeds are usually found near the typhoon center. Strong rain, on the other hand, can be experienced near dense cloud structures. Kossin\cite{Kossin18} has found out that tropical cyclone translation speed has decreased globally by $10 \%$ and in Asia by $30 \%$ over the period 1949–2016. The lower the velocity of the cyclone as a whole system, the more rain falls on one location and the higher the risk for flooding. Thus, slower translation speeds can cause even more destructive floods. Therefore, the track of a cyclone should consider both, the center and the cloud motion all around. 


This work concentrates on tracks of cyclones that form in the north-western pacific, also known as typhoons. The geographical focus lies on the Korean peninsula. In the past, Korea has suffered from numerous typhoons of different sizes. Typhoon Sarah was the strongest one. The cyclone hit the island Jeju on September 16th in 1959 during the traditional Thanksgiving festival leaving $655$ dead, $259$ missing and more than $750,000$ homeless\cite{1959}. In the younger history of the country, especially typhoons Rusa (August 2002) and Maemi (September 2003) have caused fear and death among the Korean population. Together they have been responsible for $376$ casualties and a damage of $11.5$ billion USD\cite{Kim07,Maemi}. To save lives and reduce such damages in the future, accurate forecast methods need to be established.  


Accuracy is only one criteria when talking about the prediction of natural disasters. Speed and flexibility play significant roles as well. Forecasts should be done quickly and forecast tools should be able to react immediately to sudden changes. Finally they should be inexpensive. Current predictions in South Korea are done by conducting numerical simulations on a Cray XC40 supercomputer with 139,329 CPUs. This method consumes a huge amount of computational time. The maintenance cost for such expensive hardware is also very high. An efficient forecast method is needed, that considers all criteria.

In the work of Kovordanyi and Roy (2009)\cite{Kovordanyi09} nine existing forecast techniques are listed. Subjective assessment is an evaluation of the cyclone's behavior based on large-scale changes of the flow field. In analogue forecasts, features of the cyclone are compared to all previous storms in the same region and the movement is derived from past experiences. In the third technique, the steering current is estimated by analyzing the winds at certain locations and altitudes. The statistical technique uses regression analysis, whereas the dynamical technique uses numerical modeling and simulation. Besides computer based approaches, empirical forecasting based on experiences of meteorologists are considered as a good complement to other techniques. The persistence method allows short-term predictions, but relies on the cyclone to keep its recent track. Satellite-based techniques use satellite images to make forecasts of the track and intensity of tropical cyclones based on cloud patterns. Finally, combinations of the previously mentioned approaches are defined as hybrid techniques. In this study a satellite-based technique is used.

The satellite-based approach is combined with a deep learning method. To the knowledge of the authors, Lee and Liu (2000)\cite{Lee00} firstly used satellite images of tropical cyclones for track prediction with the help of neural networks. But the images itself did not function as input data for the network. In a pre-processing step, information like the Dvorak number, the max. wind speed or the cyclone's position have been extracted from the images and fed to the network for a time-series prediction. The model has shown improvements of $30$ and $18\%$, respectively, over the existing numerical model for forecasting tropical cyclone patterns and tracks. Kovordanyi and Roy (2009)\cite{Kovordanyi09} were the first who actually used satellite images as input data for a neural network. The network successfully detected the shape of a cyclone and predicted the future movement direction. Hong et al. (2017)\cite{Hong17} utilized multi-layer neural networks to predict the position of the cyclone's eye in a single high-resolution 3D remote sensing image. The network learns the coordinates of the eye from labeled images of past data and predicts them in test images. Kordmahalleh et al. (2016)\cite{Moradi16} have used a sparse Recurrent Neural Network (RNN) to predict the trajectories of cyclones coming from the Atlantic Ocean or the Caribbean Sea, also known as Hurricanes. They have used Dynamic Time Warping (DTW), which lets the neural network learn information equally from each hurricane. Alemany et al. (2018)\cite{Alemany18} have also used a RNN to predict hurricane trajectories, but instead of assuming monotonic behavior of all hurricanes they focus on temporal and unique features of each cyclone. By learning extracted parameters, like the angle of travel or wind speed, their forecast results could achieve better accuracies than the results of Moradi Kordmahalleh et al. (2016). Zhang et al. (2018)\cite{Zhang18} have utilized a matrix neural network for predicting the trajectories of cyclones that occurred in the South Indian ocean. They state that matrix neural networks (MNN) suit better to the task of cyclone trajectory prediction than RNNs, because MNNs can preserve spatial information of cyclone tracks.  

In the past, the focus lied on predicting trajectories. But, as mentioned before, a safe warning system should be able to forecast both, cyclone centers and their area of impact in form of cloud structures. To the knowledge of the authors, a combination of satellite images and a deep learning method has not yet been used to predict both, future typhoon centers and the future cloud motion. In this study, a Generative Adversarial Network (GAN) is applied for both tasks. As input it uses chronologically ordered satellite images from the past and as output it creates images that show the typhoon one time step ahead. Predictions can be generated within seconds using a single graphics processing unit (GPU). Thus, the method is quicker and cheaper than conventional methods, like numerical simulations, where hundreds of thousands CPU are used for one simulation. Additionally, it is more flexible, since newly generated satellite images can easily be added to the input data.

A GAN is a deep learning technique used to generate samples in forms of images, music or speech. It has been introduced by Goodfellow (2014)\cite{Goodfellow14}, who successfully generated new samples to the MNIST and CIFAR-10 image datasets. The inspiration for the present study comes from a problem in fluid dynamics. Lee and You (2018)\cite{Lee18} used a GAN to successfully predict the unsteady flow field of the flow over a cylinder. Their multi-scale convolution network architecture has originally been proposed by Mathieu et al. (2015)\cite{Mathieu15} for image prediction of a video sequence.    


The paper is organized as follows. The dataset that contains the input data for the GAN is presented in section 2. The deep learning methodology is explained in section 3. This includes the architecture and the loss function of the GAN. In the fourth part the prediction results are presented and discussed. Concluding remarks finalize the study in the last part.

\section{Dataset}

\begin{figure}
  \centerline{\includegraphics[scale=0.38]{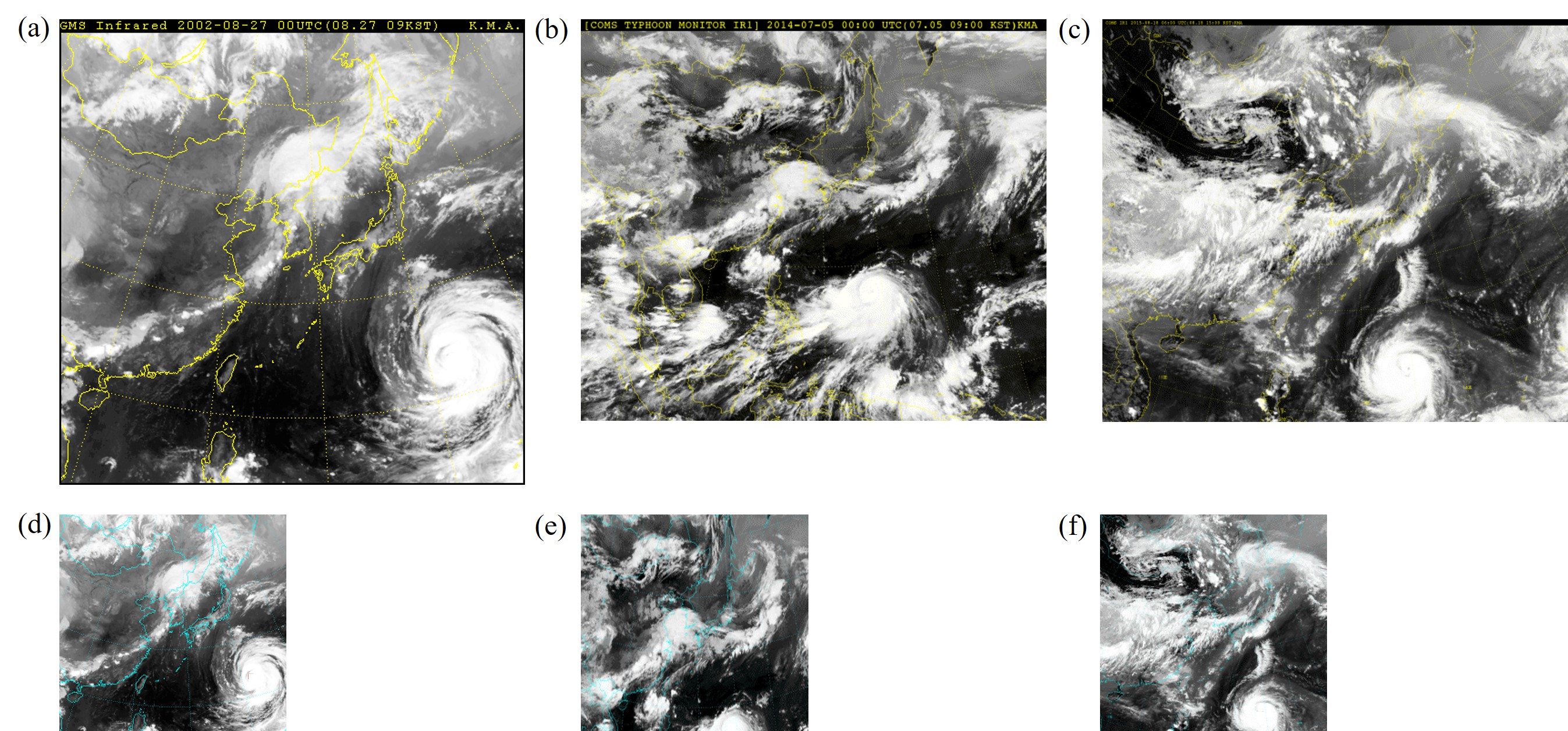}}
  \caption{Images of typhoon Rusa in August $2002$ as an example for the shots taken from $1993$ to $2010$ before {\textit(a)} and after {\textit(d)} preprocessing, images of typhoon Neoguri in July $2014$ as an example for the shots taken from $2011$ to $2014$ before {\textit(b)} and after {\textit(e)} preprocessing and images of typhoon Goni in August $2015$ as an example for the shots taken from $2015$ to $2017$ before {\textit(c)} and after {\textit(f)} preprocessing.}  
\label{fig:pre_processing}
\end{figure}

The input satellite images have been provided by the Korean Meteorological Administration (KMA)\cite{KMA}. They contain those $76$ typhoons from $1993$ till $2017$ that hit or were about to hit the Korean peninsula. Whereas Hong et al. (2017)\cite{Hong17} used satellite images directly, in this study preprocessing of the images is inevitable. During the 25 years of capture time, different satellites have been operating. Thus, the raw images have different perspectives on the Korean peninsula. Three different types of image perspectives and pixel sizes are provided (see figure~\ref{fig:pre_processing}). The network only takes images with the same pixel size and learns better if all images have the same or a similar perspective on the north-western pacific.

Figure~\ref{fig:pre_processing}(a) shows an image of typhoon Rusa from August 2002 as an example for the first perspective of the first period from $1993$ till $2010$. The images have a pixel size of $512$$\times$$512$. Initially, the images have been generated by the Geostationary Meteorological Satellite (GMS) number 4 of Japan\cite{GMS}. GMS-4 has been replaced by the GMS-5 satellite in $1996$ with a design life of five years. But the replacement failed and the GMS-5 had to operate for three more years, before the US National Oceanic $\&$ Atmospheric Administration (NOAA) helped out with its GOES-9 Satellite in $2003$. Two years later Japan could use the Multi-functional Transport Satellite (MTSAT) type 1R until $2010$. All satellites used the Visible and Infrared Spin-Scan Radiometer (VISSR) technique to obtain visible and infrared spectrum mappings of the earth and its cloud cover. The upper black bar, containing the satellite name, the date and the time written in yellow, as well as the black frame are unnecessary information for the learning phase. Furthermore, the pixel size causes memory issues in the test phase. Thus, the images have been cropped and resized to $250$$\times$$238$ pixels (see figure~\ref{fig:pre_processing}(d)).

The perspective of the images captured between $2011$ and $2014$ is presented in figure~\ref{fig:pre_processing}(b). The image with pixel size $512$$\times$$412$ shows typhoon Neoguri that formed in July 2014. It has been captured by Korea's first multi-purpose geostationary meteorological satellite, namely Communication, Ocean and Meteorological Satellite (COMS). As in the previous case, the upper black bar is cropped and the images are resized to $250$$\times$$238$ pixels (see figure~\ref{fig:pre_processing}(e)). The view on the earth in figure~\ref{fig:pre_processing}(e) is similar to the view in figure~\ref{fig:pre_processing}(d). 

An example for the raw images of the remaining data is shown in figure~\ref{fig:pre_processing}(c). The operating satellite is the same like in the previous case. The $512$$\times$$433$ pixel shot shows typhoon Goni from August 2015. Except for a different thickness of the upper black bar, the pre-processing steps match with the previous case and lead to images like in figure~\ref{fig:pre_processing}(f). 

All images have the ".png" format with three color channels (R(red)G(green)B(blue)). To improve the visibility of the country boarders, their color has been changed from yellow to blue. In total $1,628$ images are stored, with a time step size of 6 hours between the images. There are two accuracy criteria for the typhoon track prediction, a quantitative and a qualitative one. For the quantitative criterion the difference between the predicted typhoon center coordinates and the real ones are taken into consideration. In the qualitative criterion the future shape of the clouds is estimated. 

\begin{figure}
  \centerline{\includegraphics[scale=0.6]{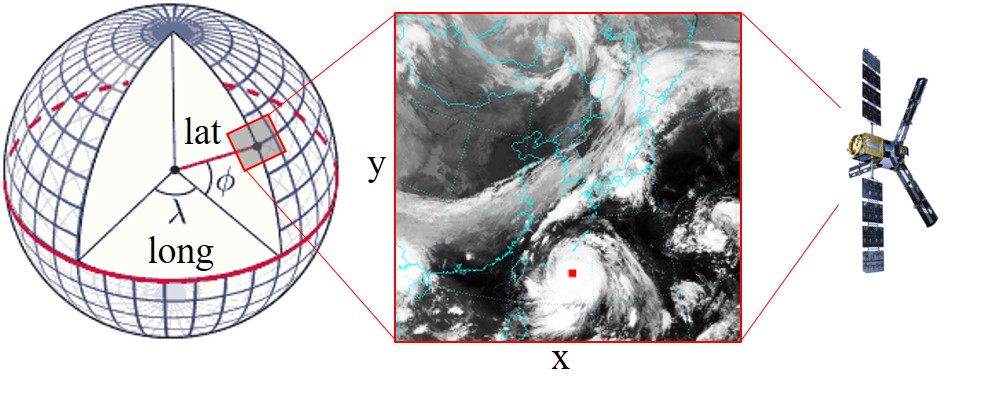}}
  \caption{Georeferencing of the typhoon center coordinates.}  
\label{fig:georeferencing}
\end{figure}

Every satellite image has been labeled with a red square at the typhoon center. The latitudinal ($\phi$) and longitudinal ($\lambda$) coordinates of the typhoon centers have been provided by the Japan Meteorological Agency (JMA)\cite{JMA}. In order to label each satellite image with its red square, the $\phi$ and $\lambda$ coordinates have to be transfered to (x,y)-pixel coordinates in the image. This process is known as georeferencing and is illustrated in figure~\ref{fig:georeferencing}.

The images are split into training and test data. The training data contain $1,389$ images of $66$ typhoons, the test data are $239$ images of $10$ typhoons. This study focuses on the development of each typhoon until it reaches the land. Typhoon centers and cloud motion after landfall are not considered. 

\section{Deep learning methodology}

The deep learning methodology is different for the training and the testing step. Before using the training data as input images to the GAN they get cropped to a total number of 5,000,000 images with a pixel size of $32$$\times$$32$. On the one hand, this reduces the memory used during the training step. On the other hand, focusing on small parts of an image allows to learn the details of the image better. The testing, on the contrary, is done with the full scale test data, containing images with $250$$\times$$238$ pixels. 

\begin{figure}
  \centerline{\includegraphics[scale=0.3]{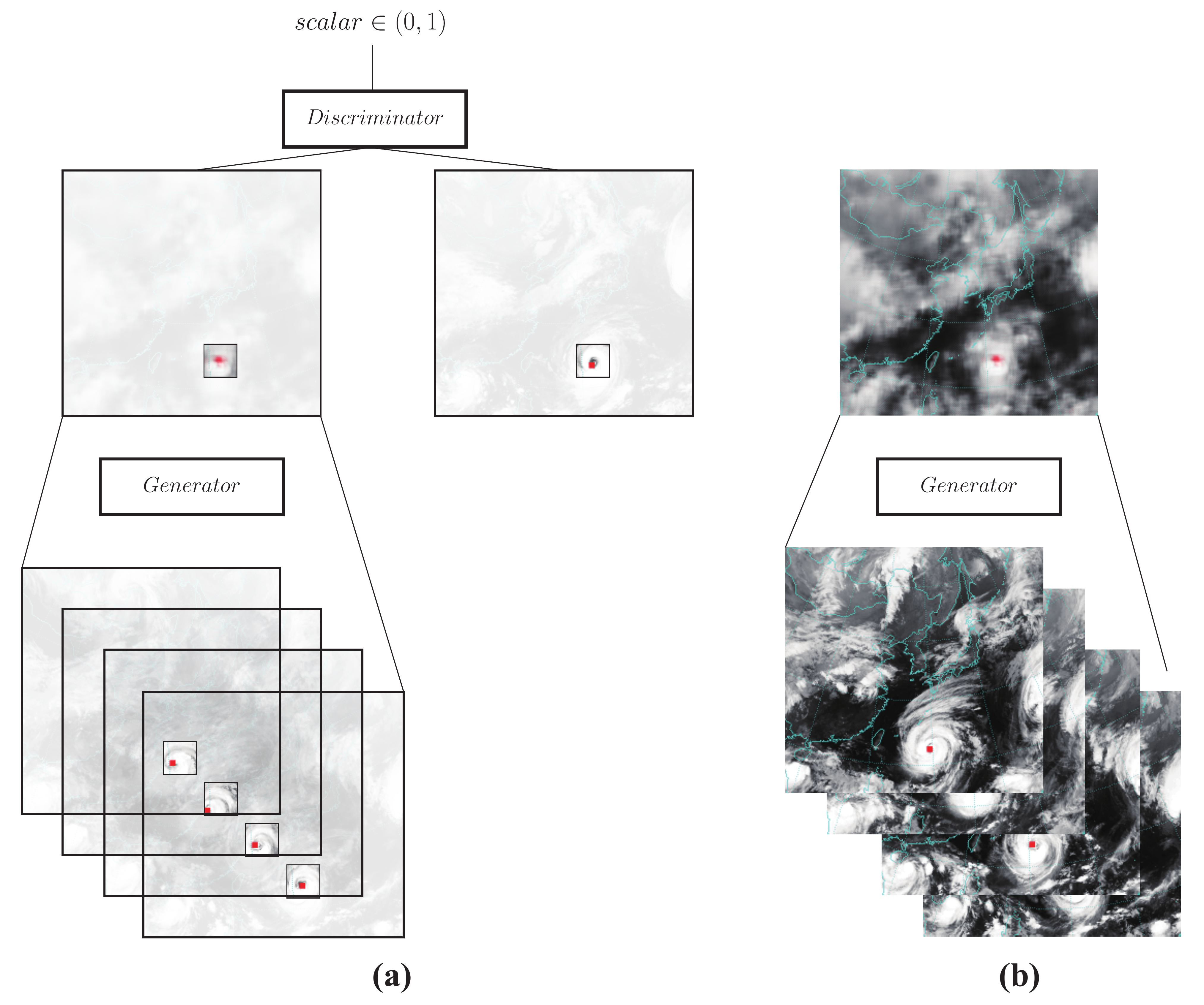}}
  \caption{Overall structure of the utilized GAN with four input images ($m=4$) for (a) training using $32 \times 32$ sized cropped images and (b) testing using $250 \times 238$ sized images.} \label{fig:network}
\end{figure}

The cropped training images are trained using variations of deep learning networks in a GAN based video modeling architecture~\cite{Mathieu15}. For each training sequence a $32$$\times$$32$ part of a typhoon satellite image at future occasion (output) is predicted based on a set of $m$ consecutive cropped satellite images from the past (input, $I$) using the GAN based deep learning network (see figure~\ref{fig:network} a). 
The configuration of the deep learning network is summarized in table~\ref{tab:network} (see Mathieu et al. (2015)\cite{Mathieu15} for the detailed algorithms).
The generator model (G) of the deep learning network is comprised of four fully convolutional networks with multi-scale architectures ($\{G_{0}, G_{1}, G_{2}, G_{3}\}$).
Each fully convolutional network ($G_{k}$, where $k=0,1,2,3$) generates predictions of the images on a $\frac{1}{2^{k}}$ times coarser image, but with a same domain size, compared to the provided ground truth images.

The discriminator model (D) of the deep learning network is comprised of a sequence of multi-scale networks ($\{ D_{0},D_{1},D_{2},D_{3}\}$) with convolution layers and fully connected layers to classify the predicted images to class 0 and the provided ground truth images to class 1. 

\begin{table}
    \centering
    \resizebox{0.9 \textwidth}{!}{%
    \begin{tabular}{c|c|c|c}
        \hline
        \multicolumn{4}{c}{Generator model} \\ \hline
        $G_3$ & $G_2$ & $G_1$ & $G_0$ \\ \hline \hline
        \multicolumn{4}{c}{Convolution layers} \\ \hline
        \multicolumn{4}{c}{(Feature map numbers)} \\ \hline
        3m 128 256 128 3 & 3(m+1) 128 256 128 3 & 3(m+1) 128 256 512 256 128 3 & 3(m+1) 128 256 512 256 128 3\\ \hline
        \multicolumn{4}{c}{(Kernel sizes)}\\ \hline
        $3\times3$, $3\times3$, $3\times3$, $3\times3$ & $5\times5$, $3\times3$, $3\times3$, $5\times5$ & $5\times5$, $3\times3$, $3\times3$, $3\times3$, $3\times3$, $5\times5$ & $7\times7$, $5\times5$, $5\times5$, $5\times5$, $5\times5$, $7\times7$\\ \hline \hline
        \multicolumn{4}{c}{Discriminator model} \\ \hline
        $D_3$ & $D_2$ & $D_1$ & $D_0$ \\ \hline \hline
        \multicolumn{4}{c}{Convolution layers} \\ \hline
        \multicolumn{4}{c}{(Feature map numbers)} \\ \hline
        3 64 & 3 64 128 128 & 3 128 256 256 & 3 128 256 512 128\\ \hline
        \multicolumn{4}{c}{(Kernel sizes)}\\ \hline
        $3\times3$ & $3\times3$, $3\times3$, $3\times3$ & $5\times5$, $5\times5$, $5\times5$ & $7\times7$, $7\times7$, $5\times5$, $5\times5$ \\ \hline \hline
        \multicolumn{4}{c}{$2 \times 2$ max pooling} \\ \hline \hline
        \multicolumn{4}{c}{Fully connected layers} \\ \hline
        \multicolumn{4}{c}{(Neuron numbers)} \\ \hline
        512 256 1 & 1024 512 1 & 1024 512 1 & 1024 512 1\\ \hline \hline
    \end{tabular}}
    \caption{Configuration of the deep learning network.}
    \label{tab:network}
\end{table}

The generator model is trained to minimize a combination of loss functions as
\begin{eqnarray}
{L}_{generator} = 1/4 \sum_{k=0}^{3}{\lambda_{l2} {L}_{2}^{k} + \lambda_{gdl} {L}_{gdl}^{k} + \lambda_{adv} {L}_{adv}^{G,k}},
\end{eqnarray}
where $\lambda_{l2} = 1$, $\lambda_{gdl} = 1$, and $\lambda_{adv} = 0.05$.
Let $G_{k}(I)$ be the predicted image from a convolutional neural network of $G_{k}$ and $\mathcal{G}_{k}(I)$ be a $\frac{1}{2^{k}}$ resized image from the provided ground truth image.
The ${L}_{2}^{k}$ loss function evaluates the explicit difference between the predicted and provided images as
\begin{eqnarray}
{L}_{2}^{k} = ||G_{k}({I}) - \mathcal{G}_{k}({I})||_{2}^{2}.
\end{eqnarray}
The $\mathcal{L}_{gdl}^{k}$ loss function compares the difference between gradients of the predicted and provided images as
\begin{eqnarray}
{L}_{gdl}^{k} =  \sum_{i=0}^{n_x-2} \sum_{j=0}^{n_y-2} \Big\{ \bigg| |\mathcal{G}_{k}({I})_{(i+1,j+1)}-\mathcal{G}_{k}({I})_{(i,j+1)}| -|G_{k}({I})_{(i+1,j+1)} - G_{k}({I})_{(i,j+1)}| \bigg| \\ \nonumber
+\bigg||\mathcal{G}_{k}({I})_{(i+1,j+1)}-\mathcal{G}_{k}({I})_{(i+1,j)}|  -|G_{k}({I})_{(i+1,j+1)} - G_{k}({I})_{(i+1,j)}| \bigg| \Big\},
\end{eqnarray}
where $n_x$ and $n_y$ are the numbers of pixels in the width and height of an image, and (i,j) are the pixel coordinates.
The $\mathcal{L}_{adv}^{G,k}$ loss function is applied to support the generator model to delude the discriminator network by generating images which are indistinguishable from the provided ground truth images as
\begin{equation}
{L}_{adv}^{G,k} = L_{bce}(D_{k}(G_{k}({I})),1),
\label{eqn: ladvg}
\end{equation}
where $L_{bce}$ is the binary cross entropy loss function, defined as
\begin{equation}
L_{bce}(a,b) = -b \log(a) - (1-b) \log(1-a),
\label{eqn: lbce}
\end{equation}
for scalars $a$ and $b$ between 0 and 1.

The discriminator model is trained to minimize a loss function as
\begin{equation}
{L}_{discriminator} = \frac{1}{4}\sum_{k=0}^{3} \left[ L_{bce}(D_{k}(\mathcal{G}_{k}({I})),1) + L_{bce}(D_{k}(G_{k}({I})),0)\right].
\label{eqn: ladvd}
\end{equation}
This supports the network not to be deluded by the generator model by extracting important features of typhoons in an unsupervised manner.

For each testing run the generator model is fed with $m$ consecutive full scale satellite images from the past and generates a full scale typhoon satellite image at future occasion (see figure~\ref{fig:network} b).

\section{Results and Discussion}
The GAN has been trained and tested on a NVIDIA Tesla K40c GPU. The ten typhoons of the test dataset are: Faye (Juli 1995), Violet (September 1996), Oliwa (September 1997), Saomai (September 2000), Rammasun (June/July 2002), Maemi (September 2003), Usagi (July/August 2007), Muifa (July/August 2011), Neoguri (June/July 2014), Malakas (September 2016). The testing for each typhoon is done in sequences. One sequence contains the chronologically ordered input images, the ground truth image and the generated image. A sequence is named with the date and time (universal time coordinated - UTC) of the corresponding ground truth image, for example: "1993072718" stands for 27th of July in 1993 at 6pm (UTC).  

\subsection{Predicting Typhoon center coordinates}

\begin{figure}
  \centerline{\includegraphics[scale=0.43]{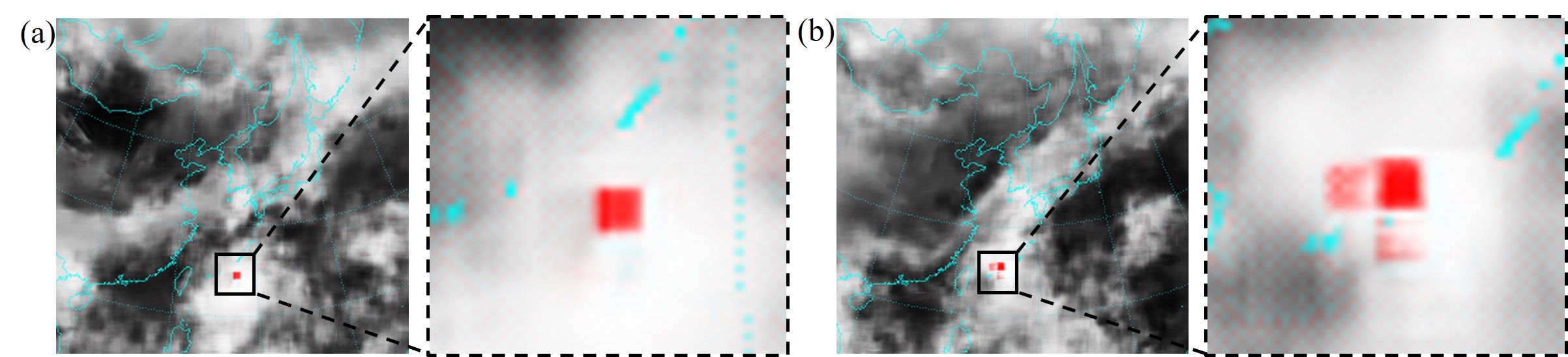}}
  \caption{(a) Clearly identifiable predicted typhoon center. (b) Three possible predicted typhoon centers.}  
\label{fig:dots}
\end{figure}

The accuracy of the typhoon center coordinate prediction is measured by comparing the labeled red square of the ground truth images with the predicted red square of the generated images. The red square in the predicted images is detected by applying a color filter on the generated images and a convolution with the size of the red square on the filtered images. The pixel with the highest value in a convolved image gives the location of the predicted typhoon center. In most of the cases, images with a clearly identifiable red square are generated, as shown in figure~\ref{fig:dots} a. In some cases, however, the GAN provides several alternatives, visualized in figure~\ref{fig:dots} b. Each possible predicted typhoon center has its own color strength. The color intensities can be interpreted as probabilities. A strong red square shows a confident prediction, a weak mark stands for a possible predicted typhoon center with a low probability. In cases like in figure~\ref{fig:dots} b, the location with the strongest color intensity is chosen.   

After identifying the (x,y)-pixel coordinates of the predicted typhoon center, they are transformed back to ($\phi$) and ($\lambda$) coordinates via georeferencing. In the current study two different kinds of errors are investigated. The first one describes the distance between the predicted coordinates ($\phi_{pred}$, $\lambda_{pred}$) and the real coordinates ($\phi_{real}$, $\lambda_{real}$), named as absolute error ($E$). E is calculated in kilometers (km) by applying the haversine formula~\cite{Mahmoud16}, with the earth radius R taken at the location of the real coordinates:

\begin{equation}
	E = 2R\arcsin{\sqrt{\sin({\frac{\phi_{pred}-\phi_{real}}{2}})^2+\cos{\phi_{real}}\cos{\phi_{pred}}\sin({\frac{\lambda_{pred}-\lambda_{real}}{2}})^2}}.
\end{equation}

However, the absolute error does not give sufficient information about the prediction quality. Tracks of fast moving typhoons, for example, are more difficult to predict than tracks of slow ones. It is therefore necessary to introduce a relative error ($E_{rel}$), that is calculated as follows:

\begin{equation}
	E_{rel}(t) = \frac{E}{2R\arcsin{\sqrt{\sin({\frac{\phi_{real}(t)-\phi_{real}(t-6h)}{2}})^2+\cos{\phi_{real}(t)}\cos{\phi_{real}(t-6h)}\sin({\frac{\lambda_{real}(t)-\lambda_{real}(t-6h)}{2}})^2}}}.
\end{equation}

This type of error considers the ratio between E and the distance that a typhoon has traveled over the last 6 hours, where $t$ stands for the time of a certain sequence.

\begin{figure}
  \centerline{\includegraphics[scale=0.4]{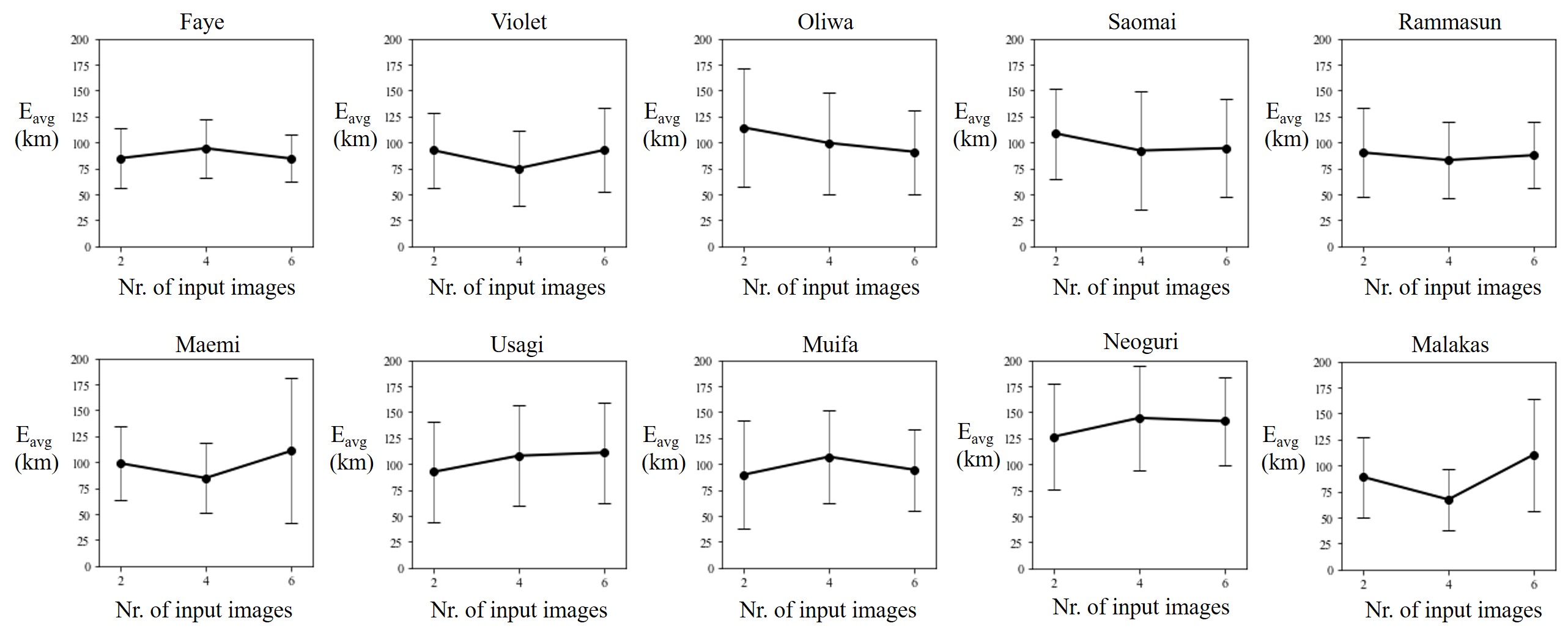}}
  \caption{Averaged absolute error and standard deviation of the absolute error for m = 2, 4 and 6 input images, applied on the ten test typhoons.}  
\label{fig:input_images}
\end{figure}

An important hyperparameter in this study is the number of input images. A high number would increases the memory usage and therefore the learning time. A low number might cause a shortage of input information for the GAN. Figure~\ref{fig:input_images} shows the averaged absolute error ($E_{avg}$) and the standard deviation of E for $m = 2$, $4$ and $6$ input images, applied on the ten test typhoons. In two cases (Faye, Oliwa) the network trained with 6 input images gives the most accurate results, in three cases (Usagi, Muifa, Neoguri) 2 input images deliver the best results, but for the majority of the cases (Violet, Saomai, Rammasun, Maemi, Malakas) 4 input images are the best choice. Relatively high standard deviations are found in two cases for a prediction with 2 input images (Oliwa, Rammsun) and in two cases for a choice of 6 input images (Maemi, Malakas). In case of 6 input images there have been three sequences where the GAN did not generate a red square and a prediction has not been possible. 

\begin{figure}
  \centerline{\includegraphics[scale=0.55]{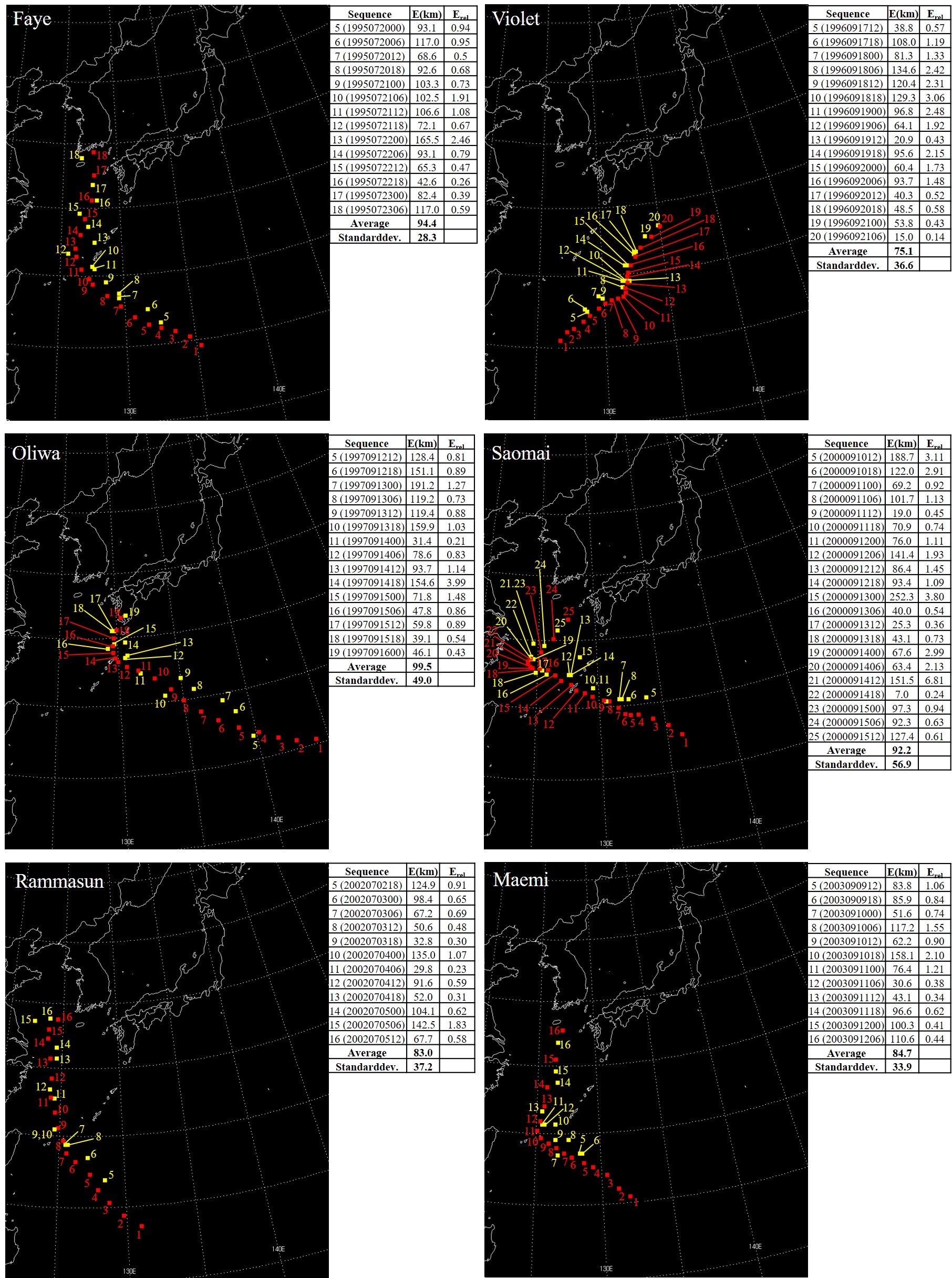}}
  \caption{Real (red) and predicted (yellow) trajectories, absolute errors and relative errors using 4 input images for the typhoons: Paye, Violet, Oliwa, Saomai, Rammasun and Maemi.}  
\label{fig:results_4_input}
\end{figure}

\begin{figure}
  \centerline{\includegraphics[scale=0.55]{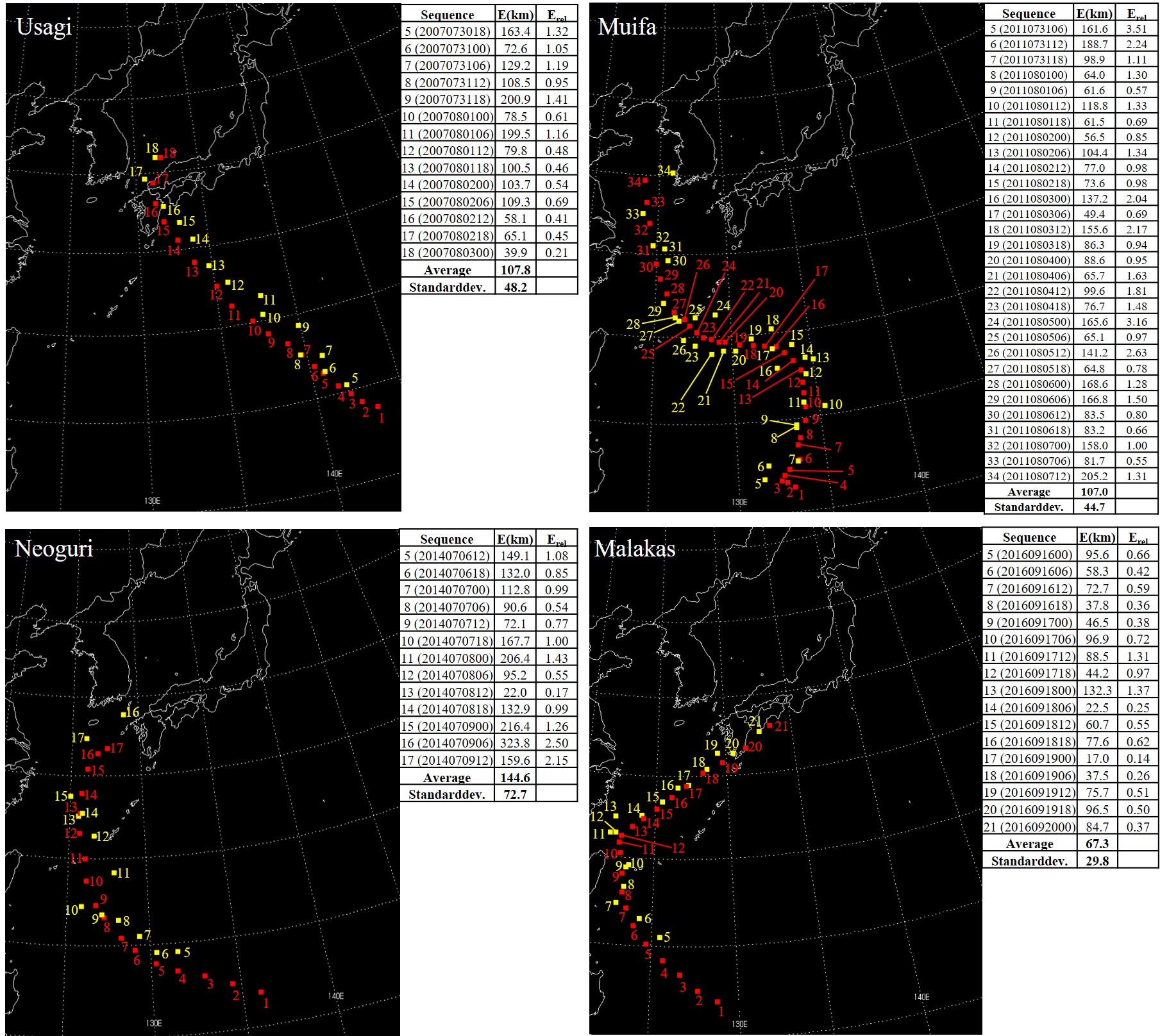}}
  \caption{Real (red) and predicted (yellow) trajectories, absolute errors and relative errors using 4 input images for the typhoons: Usagi, Muifa, Neoguri, Malakas.}  
\label{fig:results_4_input_1}
\end{figure}

Based on the previous findings, it is recommended to use 4 input images. The results with 4 input images in forms of real (red) and predicted (yellow) trajectories, $E$ and $E_{rel}$ for each sequence are shown in figures~\ref{fig:results_4_input} and~\ref{fig:results_4_input_1}. 

Typhoon Faye reached the island Jeju on July 23rd in 1995. The predictions have the lowest fluctuations among all cases. The best prediction results are achieved when the typhoon approaches Jeju. The largest relative errors are found at sequences 10, 11 and 13, when the cyclone turns northward. Typhoon Violet was expected to become a serious threat to the Korean peninsula in its initial stage, but it changed its course towards Japan in the middle of September 1996. Similar to Faye, when Violet changes its course at sequences 8-11, the predictions become less accurate. When it approaches the Japanese mainland, however, the GAN predicts the typhoon center accurately with absolute errors of less than 50 km. Cyclone Oliwa caused losses at off shore South Korea in September 1997. The prediction quality is low when moving above the open water at sequence 7, and when turning westward at sequence 10 and northward at sequences 13-15. Good results are achieved between the course changes and at the final sequences. The predictions for typhoon Saomai show the potential of the GAN with an absolute error of less than $20$ km at sequence 9 or less than $10$ km at sequence 22. However, the predicted centers show increased gaps above the open water at sequences 5 and 6. The cyclone had two sudden course changes. At sequence 15 it turned towards the Chinese mainland and was supposed to hit the area around Shanghai, when it suddenly changed its course again at sequence 21, moving towards South Korea and Japan. Both course changes cause problems to the prediction accuracy. Typhoon Rammasun did not have significant course changes in summer 2002. Only a minor course correction at sequence 10 results in a slightly increased error. The highest error is found before landfall at sequence 15. The predictions for Maemi, one of the strongest and most dangerous typhoons that ever hit the Koren peninsula, again show a weakness when getting deflected northwards at sequences 8-11. No issues with changing course, but problems above the open water are observed for typhoon Usagi, especially at sequences 5,9 and 11. Once the cyclone approaches Japan and South Korea, the errors become more and more accurate. In August 2011 the dragon shaped typhoon Muifa found its way to the Korean mainland. Again, it is eye-catching how the GAN struggles initially at sequences 5-7, as well as with sudden course changes at sequences 16-18 or 24-26. The highest absolute error is found at sequence 34 right before the typhoon hits the Korean peninsula. A similar observation is made for the predicted centers of typhoon Neoguri, that reveal high errors before landfall. The youngest typhoon among the ten test cases, Malakas, shows the best averaged error. Except for the slight increase in relative error at sequences 11-13, it shows the potential of GAN in predicting typhoon tracks.

Table~\ref{tab:frequency_distribution} provides the frequency distribution of $E$ for all sequences. In $74.5 \%$ of the cases the error is less than or equal to $120$ km. A combined $42.4 \%$ have accuracies of less than or equal to $80$ km. The remaining errors are found above $120$ km. 

The relative error provides information about how close the predicted typhoon center is to the real one. Most of the cases with high relative errors deal with three major challenges. Firstly, it can be difficult to predict the motion above open water, when the typhoons have many possible paths to take and their path is not limited by land. Secondly, sudden changes in direction can cause increased relative errors. Finally, sometimes the relative error in sequences right before landfall is difficult to predict. All problems seem to deal with a lack of information for the GAN. The predictions are made based on satellite images and typhoon center coordinates. But, as described in the introduction of this work, typhoons are a complex natural phenomenon depending on the temperature and pressure, as well as the wind velocities in their surroundings. Furthermore, the clouds right before landfall begin to dissipate and the satellite images lose valuable information. In order to predict open water movement, sudden course changes or the motion before landfall accurately, it is recommended to add more information to the input data, like for example the sea surface temperature, the surface pressure or the velocity field at a certain height.  

\begin{table}[ht]
\centering
\begin{tabular}{|l|l|l|}
\hline
E (km) & Sequences & Percentage (\%) \\
\hline
0-40 & 18 & 10.9 \\
\hline
41-80 & 52 & 31.5 \\
\hline
81-120 & 53 & 32.1 \\
\hline
$\geq$ 121 & 42 & 25.5 \\
\hline
\end{tabular}
\caption{Error distribution for the prediction of all sequences using 4 input images.}
\label{tab:frequency_distribution}
\end{table}

\subsection{Predicting cloud motion}

The cloud structure on the satellite images is not represented in physical units. Thus, rather than measuring the accuracy of the cloud motion prediction quantitatively, this study focuses on predicting trends in cloud motions qualitatively. 

\begin{figure}
  \centerline{\includegraphics[scale=0.4]{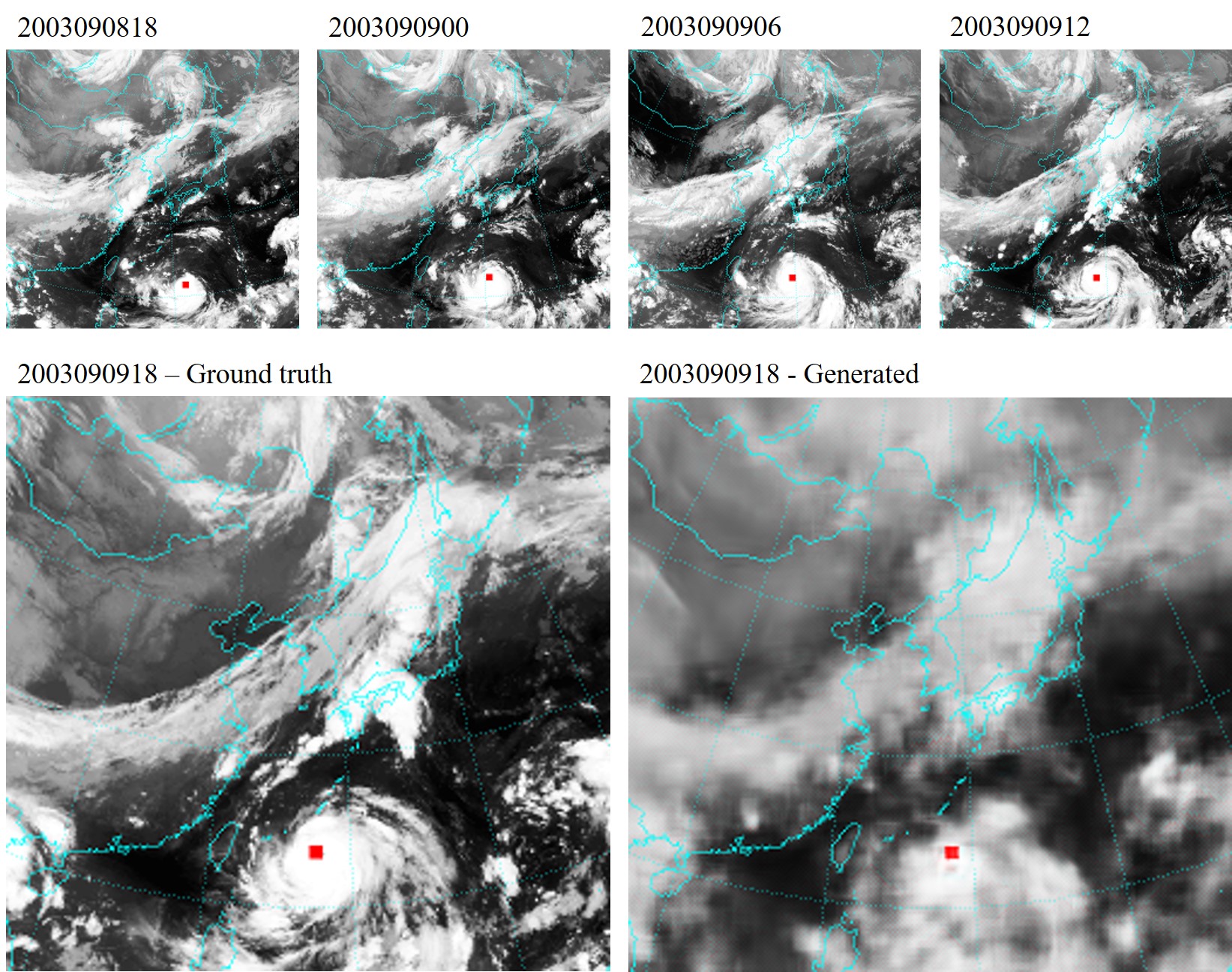}}
  \caption{Prediction of sequence 6 (2003090918) for typhoon Maemi using 4 input images.}  
\label{fig:blurry}
\end{figure}

As an example for a generated image, figure~\ref{fig:blurry} illustrates the prediction for sequence 6 of typhoon Maemi. The generated image does not have the same sharpness than the ground truth image, instead it is blurry. It cannot clearly show details, like the spinning motion of the typhoon. Blurriness is a well known challenge in video prediction tasks that is mainly influenced by the gdl part of the loss function. However, although the generated image suffers from a certain degree of blurriness, the main structure of the clouds is still detectable. 

\begin{figure}
  \centerline{\includegraphics[scale=0.4]{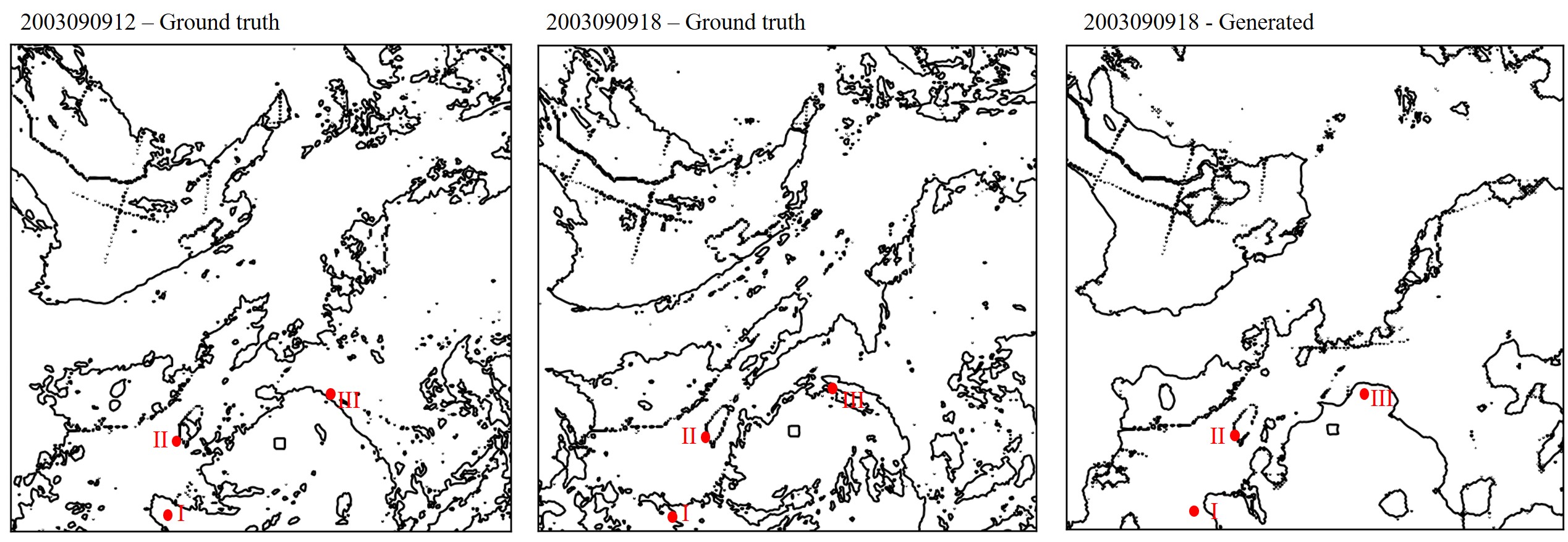}}
  \caption{Contour plots of ground truth images of sequences 5 (2003090912) and 6 (2003090918), as well as the generated image of sequence 6 (2003090918) for typhoon Maemi using 4 input images.}  
\label{fig:cloud_motion}
\end{figure}

A way to deal with the blurry images and extract information about trends of cloud motion is to look at the clouds in form of contour plots. This is shown in figure~\ref{fig:cloud_motion} for the same sequence that has been presented in figure~\ref{fig:blurry}. Closer looks at the ground truth image one time step afore at sequence 5 and the generated image and ground truth image at sequence 6 give clues about trends that the GAN is able to predict. The reduction in cloud density in the northern part of the Philippines, marked with $I$, the dissipation of cloud structures in Taiwan, marked with $II$, or the eastward expansion of the clouds at the open sea, marked with $III$, are only three examples. The overall cloud structures away from the typhoon show a good conformity between the ground truth image and the generated one at sequence 6.

\section{Conclusion}

The application of a deep learning method for typhoon track prediction in forms of typhoon center coordinates and cloud structures has been explored. $42.4 \%$ of all typhoon center predictions have absolute errors of less than $80$ km, $32.1 \%$ lie within a range of $80$ - $120$ km and the remaining $25.5 \%$ have accuracies above $120$ km. Taking a look at the relative errors, three types of challenges have been revealed: the movement of a typhoon at open sea far away from land, the forecast of sudden course changes and the motion right before landfall. Furthermore, it has been shown that the GAN is able to predict overall trends in cloud motion. 

To improve the prediction accuracy, the next step will be adding physical information to the input data, like the sea surface temperature, the surface pressure or the velocity field in a certain height. Just learning from satellite images that show the cloud structure and the typhoon center is a good starting point but not enough for learning the complex phenomena that are responsible for the creation and motion of typhoons. 

\bibliography{sample}

\begin{thebibliography}{10}
\urlstyle{rm}
\expandafter\ifx\csname url\endcsname\relax
  \def\url#1{\texttt{#1}}\fi
\expandafter\ifx\csname urlprefix\endcsname\relax\def\urlprefix{URL }\fi
\expandafter\ifx\csname doiprefix\endcsname\relax\def\doiprefix{DOI: }\fi
\providecommand{\bibinfo}[2]{#2}
\providecommand{\eprint}[2][]{\url{#2}}

\bibitem{spiegel18}
\bibinfo{author}{{C. Schrader}}.
\newblock \bibinfo{title}{Hurrikane und taifune: Wirbelstürme bewegen sich
  immer langsamer} (\bibinfo{year}{2018}).
\newblock
  \bibinfo{note}{\url{http://www.spiegel.de/wissenschaft/natur/hurrikane-und-taifune-wirbelstuerme-bewegen-sich-immer-langsamer-a-1211499.html},
  Last accessed on 2018-08-09}.

\bibitem{Kossin18}
\bibinfo{author}{Kossin, J.}
\newblock \bibinfo{journal}{\bibinfo{title}{A global slowdown of
  tropical-cyclone translation speed}}.
\newblock {\emph{\JournalTitle{nature}}} \textbf{\bibinfo{volume}{558}},
  \bibinfo{pages}{104--107} (\bibinfo{year}{2018}).

\bibitem{1959}
\bibinfo{author}{Tilden, C.}
\newblock \bibinfo{journal}{\bibinfo{title}{Annual typhhon report}}.
\newblock {\emph{\JournalTitle{U.S. fleet weather central / Joint typhoon
  warning center}}}  (\bibinfo{year}{1959}).

\bibitem{Kim07}
\bibinfo{author}{Kim, A.}, \bibinfo{author}{Tachikawa, Y.} \&
  \bibinfo{author}{Takara, K.}
\newblock \bibinfo{journal}{\bibinfo{title}{Recent flood disasters and progress
  of disaster management system in korea}}.
\newblock {\emph{\JournalTitle{Annuals of Disas. Prev. Res. Inst., Kyoto
  University}}} \textbf{\bibinfo{volume}{50 B}} (\bibinfo{year}{2007}).

\bibitem{Maemi}
\bibinfo{author}{{National emergency management}}.
\newblock \bibinfo{title}{Maemi - disaster reports} (\bibinfo{year}{2014}).
\newblock
  \bibinfo{note}{\url{https://web.archive.org/web/20131029185710/http://eng.nema.go.kr/sub/cms3/3_2.asp},
  Last accessed on 2018-08-09}.

\bibitem{Kovordanyi09}
\bibinfo{author}{Kovordanyi, R.} \& \bibinfo{author}{Roy, C.}
\newblock \bibinfo{title}{Cyclone track forecasting based on satellite images
  using artificial neural networks}.
\newblock \bibinfo{type}{Tech. Rep.} (\bibinfo{year}{2009}).

\bibitem{Lee00}
\bibinfo{author}{Lee, R. S.~T.} \& \bibinfo{author}{Liu, J. N.~K.}
\newblock \bibinfo{journal}{\bibinfo{title}{Tropical cyclone identification and
  tracking system using integrated neural oscillatory elastic graph matching
  and hybrid rbf network track mining techniques}}.
\newblock {\emph{\JournalTitle{IEEE Transactions on Neural Networks}}}
  \textbf{\bibinfo{volume}{11}}, \bibinfo{pages}{680--689}
  (\bibinfo{year}{2000}).

\bibitem{Hong17}
\bibinfo{author}{Hong, S.}, \bibinfo{author}{Kim, S.}, \bibinfo{author}{Joh,
  M.} \& \bibinfo{author}{Song, A.}
\newblock \bibinfo{journal}{\bibinfo{title}{Globenet: convolutional neural
  networks for typhoon eye tracking from remote sensing imagery}}.
\newblock {\emph{\JournalTitle{arXiv:1708.03417v1}}}  (\bibinfo{year}{2017}).

\bibitem{Moradi16}
\bibinfo{author}{Kordmahalleh, M.~M.}, \bibinfo{author}{Sefidmazgi, M.~G.} \&
  \bibinfo{author}{Homaifar, A.}
\newblock \bibinfo{journal}{\bibinfo{title}{A sparse recurrent neural network
  for trajectory prediction of atlantic hurricanes}}.
\newblock {\emph{\JournalTitle{In proceedings of the genetic and evolutionary
  computation conference.}}} \bibinfo{pages}{957--964} (\bibinfo{year}{2016}).

\bibitem{Alemany18}
\bibinfo{author}{Alemany, S.}, \bibinfo{author}{Beltran, J.},
  \bibinfo{author}{Perez, A.} \& \bibinfo{author}{Ganzfried, S.}
\newblock \bibinfo{journal}{\bibinfo{title}{Predicting hurricane trajectories
  using a recurrent neural network}}.
\newblock {\emph{\JournalTitle{arXiv:1802.02548v2}}}  (\bibinfo{year}{2018}).

\bibitem{Zhang18}
\bibinfo{author}{Zhang, Y.}, \bibinfo{author}{Chandra, R.} \&
  \bibinfo{author}{Gao, J.}
\newblock \bibinfo{journal}{\bibinfo{title}{Cyclone track prediction with
  matrix neural networks}}.
\newblock {\emph{\JournalTitle{Conference paper}}}  (\bibinfo{year}{2018}).

\bibitem{Goodfellow14}
\bibinfo{author}{Goodfellow, I.} \emph{et~al.}
\newblock \bibinfo{journal}{\bibinfo{title}{Generative adversarial networks}}.
\newblock {\emph{\JournalTitle{arXiv:1406.2661v1}}}  (\bibinfo{year}{2014}).

\bibitem{Lee18}
\bibinfo{author}{Lee, S.} \& \bibinfo{author}{You, D.}
\newblock \bibinfo{journal}{\bibinfo{title}{Data-driven prediction of unsteady
  flow fields over a circular cylinder using deep learning}}.
\newblock {\emph{\JournalTitle{arXiv preprint arXiv:1804.06076v1}}}
  (\bibinfo{year}{2018}).

\bibitem{Mathieu15}
\bibinfo{author}{Mathieu, M.}, \bibinfo{author}{Couprie, C.} \&
  \bibinfo{author}{LeCun, Y.}
\newblock \bibinfo{journal}{\bibinfo{title}{Deep multi-scale video prediction
  beyond mean square error}}.
\newblock {\emph{\JournalTitle{arXiv preprint:1511.05440}}}
  (\bibinfo{year}{2015}).

\bibitem{KMA}
\bibinfo{author}{{Korean Meteorological Administration (KMA)}}
  (\bibinfo{year}{2018}).
\newblock \bibinfo{note}{\url{https://web.kma.go.kr/eng/}, Last accessed on
  2018-08-09}.

\bibitem{GMS}
\bibinfo{author}{{Sharing Earth Observation Resources(eo)}}.
\newblock \bibinfo{title}{Gms (geostationary meteorological satellite) series
  of japan} (\bibinfo{year}{2018}).
\newblock
  \bibinfo{note}{\url{https://directory.eoportal.org/web/eoportal/satellite-missions/g/gms},
  Last accessed on 2018-08-09}.

\bibitem{JMA}
\bibinfo{author}{{Japan Meteorological Agency (JMA)}} (\bibinfo{year}{2018}).
\newblock \bibinfo{note}{\url{https://www.jma.go.jp/jma/indexe.html}, Last
  accessed on 2018-08-09}.

\bibitem{Mahmoud16}
\bibinfo{author}{Mahmoud, H.} \& \bibinfo{author}{Akkari, N.}
\newblock \bibinfo{journal}{\bibinfo{title}{Shortest path calculation: a
  comparative study for location-based recommender system}}.
\newblock {\emph{\JournalTitle{2016 World Symposium on Computer Applications
  and Research (WSCAR)}}}  (\bibinfo{year}{2016}).

\end{thebibliography}

\section*{Acknowledgements}

This work was supported by the National Research Foundation of Korea (NRF) under the Project Number NRF-2017R1E1A1A03070514. Computational resources have been provided by the Flow Physics and Engineering Laboratory at Pohang University of Science and Technology.

\section*{Author contributions statement}

All authors proposed the study. M.R. processed the images and analyzed the data. S.L. provided the source code. All authors discussed the results and participated in completing the manuscript. 

\section*{Additional information}

The authors declare no competing financial interests.

\end{document}